\documentstyle[aps,epsf,multicol,psfig]{revtex}
\begin{document}
\draft
\title{Theory of electron-phonon dynamics in insulating nanoparticles}
\author{Michael R. Geller,$^1$ W. M. Dennis,$^1$ Vadim A. Markel,$^2$ Kelly R.
Patton,$^1$ Daniel T. Simon,$^3$ and Ho-Soon Yang$^4$}
\address{$^1$Department of Physics and Astronomy, University of Georgia, 
Athens, Georgia 30602-2451}
\address{$^2$Department of Electrical Engineering, Washington University,
St. Louis, Missouri 63130-4899}
\address{$^3$Department of Physics, University of California,
Santa Cruz, California 95064}
\address{$^4$Argonne National Laboratory, Materials Science Division, 9700 S. 
Cass Avenue, Argonne, Illinois 60439}
\date{July 11, 2001}
\maketitle

\begin{abstract}
We discuss the rich vibrational dynamics of nanometer-scale semiconducting and
insulating crystals as probed by localized electronic impurity states, with an 
emphasis on nanoparticles that are only weakly coupled to their environment.  
Two principal regimes of electron-phonon dynamics are distinguished, and a 
brief survey of vibrational-mode broadening mechanisms is presented. Recent
work on the effects of mechanical interaction with the environment is 
discussed.
\end{abstract}

\vskip 0.05in



\begin{multicols}{2}

\newpage

\section{introduction}

Recently there has been considerable interest in nanoparticles and nanocrystals
because of their unique physical properties and for their potential use in 
nanotechnology\cite{Alivisatos}. Like quantum dots, nanocrystals made from 
semiconductors and insulators have a discrete electronic spectrum, but unlike 
quantum dots they also have a nearly discrete vibrational spectrum with a 
suppressed phonon density-of-states (DOS) at low energy. In particular, an 
isolated spherical nanoparticle of diameter $d$ cannot support internal 
vibrations at frequencies less than the so-called Lamb mode frequency, about 
$2\pi v_{\rm t}/d$, where $v_{\rm t}$ is the bulk transverse sound velocity
\cite{Lamb}. Any property of a nanoparticle that depends on the low-frequency 
vibrational spectrum, such as its low-temperature thermodynamic properties or 
low-energy electronic relaxation rate, will be different than in bulk 
crystals. This will be especially dramatic for nanoparticles---for example, 
prepared in powder form---that are only weakly coupled to their environments.

One way to probe the vibrational spectrum of a nanoparticle is to optically
excite an electron-hole pair and study the intraband electronic energy 
relaxation prior to radiative recombination \cite{Woggon and Gaponenko}. 
However, the excitonic states, being only weakly localized, will suffer 
significant quantum-confinement effects in the nanoparticle, making comparison
with bulk relaxation rather indirect. An alternative probe of the vibrational 
spectrum is provided by well-localized electronic impurity states in a doped 
nanocrystal. The impurity states can be used to probe both energy relaxation 
by phonon emission \cite{Yang etal} and phonon-induced dephasing 
\cite{Meltzer and Hong}. In these cases, the difference between the 
nanocrystal and bulk behavior is almost entirely a consequence of their 
differing vibrational modes.

Yang {\it et al.} \cite{Yang etal} measured the DOS of ${\rm Y_2 O_3}$ 
nanoparticles, in powder form, with a size distribution ranging from 7 to 23 
nm in diameter. ${\rm Y_2 O_3}$ nanoparticles 15 nm in diameter cannot support
phonons with energies below about 10 ${\rm cm^{-1}}$. The phonon DOS was 
obtained by measuring the nonradiative lifetime of an excited electronic state
of Eu$^{3+}$, and at 3 ${\rm cm^{-1}}$ was found to be about 100 times smaller
than that of a bulk ${\rm Y_2 O_3}$ crystal.

In this paper we discuss recent theoretical work \cite{Simon and Geller,Markel
and Geller,Patton and Geller,Yang and Geller} motivated by the experiment of 
Yang {\it et al.} \cite{Yang etal}. After discussing the two principal regimes
of electron-phonon dynamics, a brief survey of vibrational-mode broadening 
mechanisms is presented, and we discuss new work on the phonon DOS in a 
nanoparticle mechanically coupled to its surroundings.

\section{relaxational versus vibronic dynamics}

The simple discussion presented above suggests that phonon emission should be 
considerably suppressed below frequencies of the order of $2\pi v_{\rm t}/d$. 
However, this is only partially correct, as can be seen by considering the 
limit of a localized electronic impurity state in a {\it completely} isolated 
nanoparticle. In this case energy is exchanged between the electronic and 
vibrational degrees of freedom in a oscillatory manner, and no relaxation 
would be observed. Indeed, an impurity state in an isolated nanoparticle may 
be regarded as a phonon analog of an atom in a cavity, which is known to 
exhibit Rabi-like population dynamics \cite{Scully and Zubairy}.

We are therefore led to the conclusion that there will be two distinct regimes
characterizing the nature of low-energy electron-phonon dynamics in a 
semiconducting or insulating nanoparticle. In what we shall refer to as the 
{\it relaxational regime}, an excited-state electron population decays 
approximately exponentially in time with a rate correctly given by Fermi's 
golden rule. In the {\it vibronic regime}, however, the population oscillates 
for some number of cycles before decaying. The crossover between these regimes
occurs when the width of the broadened Lamb mode becomes equal to the 
electron-phonon interaction strength.

It is tempting to conclude that it would be straightforward to determine
experimentally whether a system of nanoparticles was in the relaxational or 
vibronic regimes, but the effects of ensemble-averaging complicate the 
matter\cite{Simon and Geller}. We have found that at short times the 
ensemble-averaged excited-state population oscillates but has a decaying 
envelope. At long times, however, the oscillations become purely sinusoidal 
about a ``plateau'' population, with a frequency determined by the 
electron-phonon interaction strength, and with an envelope that decays 
algebraically as $t^{-{1/2}}$. Whether one can observe luminescence from the 
plateau region depends on the available experimental resolution.

\section{intrinsic phonon-broadening mechanisms}

Phonon broadening can come from mechanisms operative within a nanoparticle,
or from its interaction with the environment. The former tends to be more
important at high energies and the latter at low energies.

For example, anharmonic interaction, which is always present, will cause 
the modes to broaden, leading (at finite temperature) to a DOS at energies 
below $2\pi v_{\rm t}/d$. However, anharmonicity is ineffective at low energy 
and a calculation of the anharmonic broadening showed that the resulting DOS 
is much smaller than that observed\cite{Markel and Geller}. It is clear from
energy conservation that a low-temperature DOS below  $2\pi v_{\rm t}/d$ must
come from interaction of the nanoparticle with its surroundings.

\section{extrinsic phonon-broadening mechanisms}

An important extrinsic broadening mechanism results from the fact that the
nanoparticles of Ref.~\cite{Yang etal} are not isolated, but rather are in 
contact with each other or some support structure. This contact enables the 
nanoparticles to couple to an environment with a continuous spectrum at low 
energy. Preliminary work suggests that it might be is this mechanical coupling
to the environment that broadens the vibrational modes enough to explain the 
observed DOS. 

We consider a single isotropic elastic sphere of diameter 10 nm, representing 
the nanoparticle, connected to a semi-infinite isotropic elastic continuum 
lying in the $x y$ plane and extending to infinity in the negative $z$ 
direction. For simplicity, we take the substrate and the nanoparticle to be 
made of the same material. We model the contact between the two by a weak 
harmonic spring, corresponding to the situation where the nanoparticle and
substrate are connected by only a few atomic bonds or by a small ``neck'' of 
material. The Hamiltonian for the system is
\begin{equation}
H = \sum_J \omega_J a_{J}^{\dagger} a_J + \sum_I \omega_I b_{I}^{\dagger}b_{I} 
+ \textstyle{K \over 2} \big[u^{z}_{\rm n}({\bf r}_0) - u^{z}_{\rm s}
({\bf r}_0) \big]^2,
\end{equation}
where the $a_J$ and $a_J^{\dag}$ are annihilation and creation operators for 
phonons in the nanoparticle, and the index $J$ runs over all the modes of the 
nanoparticle. The $b_I$, $b_I^{\dag}$, and $I$ correspond to the substrate 
phonons. The spring constant $K$ is taken to be of the order of an atomic bond
strength of the material. ${\bf u}_{\rm n}({\bf r})$ and ${\bf u}_{\rm s}({\bf
r})$ are the phonon displacement field of the nanoparticle and the substrate,
respectively, and ${\bf r}_0$ is the point of connection. 

The phonon DOS is obtained by calculating the nanoparticle phonon propagator
to leading order in $K$ and using the quasiparticle-pole approximation. This
method is equivalent to using Fermi's golden rule to calculate phonon lifetimes
and assuming Lorentzian lineshapes. At 3 ${\rm cm}^{-1}$ the DOS is found to 
be $5.8 \times 10^{-6} \ {\rm states \ per \ cm^{-1}}$, or about $3.2 \times 
10^{-4}$ times the bulk DOS. Experimentally, the ratio of nanoparticle to bulk
DOS was found to be approximately $7.4 \times 10^{-3}$, about 20 times larger 
than our result. This is a reasonable result considering the simplicity of our
model. However, the quasiparticle-pole approximation is inaccurate at low
energies, where the deviation from Lorentzian line shapes becomes important. 
But the order-of-magnitude agreement does suggest that we have correctly
identified the relevant broadening mechanism in these nanoparticles. An 
accurate calculation of the nanoparticle DOS, based on the solution of the 
Dyson equation, is in progress.

We have also shown that fictitious forces acting on an electron in an 
oscillating nanoparticle can induce relaxation under the right circumstances
\cite{Yang and Geller}, but we do not believe that this mechanism is operative
here. A broadening mechanism that we have not yet considered is the effect of 
adsorbed molecules or ``dirt'' on the outside of the nanoparticle, which may 
be important. 

\acknowledgements

This work was supported by NSF Grants DMR-0093217 and DMR-9871864, and by a 
Research Innovation Award and Cottrell Scholars Award from the Research 
Corporation. It is a pleasure to thank Steve Lewis and Richard Meltzer for 
useful discussions.

\end{multicols}

\end{document}